# A Transfer Learning Approach for Microstructure Reconstruction and Structure-property Predictions

Xiaolin Li[1], Yichi Zhang[2], He Zhao[2], Craig Burkhart[3], L Catherine Brinson[2,4,5], Wei Chen[2,*]

[1]Theoretical and Applied Mechanics Program, Northwestern University, Evanston, IL, 60208
[2]Department of Mechanical Engineering, Northwestern University, Evanston, IL, 60208
[3]Global Materials Science Division, The Goodyear Tire and Rubber Company, Akron, OH 44305, USA
[4]Department of Materials Science and Engineering, Northwestern University, Evanston IL, 60208
[5]currently at Department of Mechanical Engineering and Materials Science, Duke University, Durham NC, 27708
*corresponding author, contact: weichen@northwestern.edu

## Abstract

Stochastic microstructure reconstruction has become an indispensable part of computational materials science, but ongoing developments are specific to particular material systems. In this paper, we address this generality problem by presenting a transfer learning-based approach for microstructure reconstruction and structure-property predictions that is applicable to a wide range of material systems. The proposed approach incorporates an encoder-decoder process and feature-matching optimization using a deep convolutional network. For microstructure reconstruction, model pruning is implemented in order to study the correlation between the microstructural features and hierarchical layers within the deep convolutional network. Knowledge obtained in model pruning is then leveraged in the development of a structure-property predictive model to determine the network architecture and initialization conditions. The generality of the approach is demonstrated numerically for a wide range of material microstructures with geometrical characteristics of varying complexity. Unlike previous approaches that only apply to specific material systems or require a significant amount of prior knowledge in model selection and hyper-parameter tuning, the present approach provides an off-the-shelf solution to handle complex microstructures, and has the potential of expediting the discovery of new materials.

## Key words:
Microstructure reconstruction, structure-property prediction, deep convolutional networks, transfer learning, unsupervised learning.

## Introduction

Under the Materials Genome Initiative (MGI)[1], materials informatics has become a revolutionary interdisciplinary research area fundamentally changing the methods to discover and develop advanced materials. In past success using materials informatics, *stochastic microstructure reconstruction* – the process of generating one or a few microstructures with morphology embodied by a set of statistically equivalent characteristics – has demonstrated its significance in both processing-structure-property modeling[2-4] and computational materials design[5,6]. Therefore, the prescription of these microstructural characteristics is crucial in determining the effectiveness of microstructure reconstruction. Existing approaches for quantifying microstructure characteristics can be roughly classified into three major categories, i.e., approaches that are statistical modeling-based, visual features-based and deep learning-based.

Statistical modeling-based approaches employ statistical models or attributes (e.g., mean particle size) to quantify microstructure morphology or features. These methods are widely used, yet their application to microstructure reconstruction is often limited to certain types of material systems and cannot be generalized. For instance, while N-point correlation functions[7] are theoretically sound for microstructure characterization, it is computationally intractable to use high-order correlation functions (e.g., 3-point correlation and above) for microstructure reconstruction. The physical descriptor-based approach[8] is often limited to characterizing and reconstructing microstructures with regular geometries (e.g., spherical clusters) but is not applicable to material systems with irregular inclusions (e.g., ceramics or copolymer blends). Another set of examples are statistical models are approaches based on Gaussian Random Fields[9] and Markovian Random Fields[10,11]. A limitation of these approaches is the assumption that locally invariant properties always hold throughout microstructures, which is not always the case.

In the last decade, visual features used for object classification or face detection in the field of computer vision have been utilized by material scientists to characterize microstructures and to study structure-property relationships. For instance, DeCost et al.[12] used bag of visual features such as Scale-Invariant Feature Transform (SIFT) to collect a "visual dictionary" for describing and classifying microstructures. Chowdhury et al.[13] utilized visual features such as histogram of oriented gradients (HoG) and local binary patterns (LBP) to distinguish between micrographs that depict dendritic morphologies from those that do not contain similar microstructural features. Despite these successes, the use of visual features-based approaches in microstructure reconstruction is unexplored and potentially limited because these visual features are essentially low-order abstractions of microstructures, thereby rendering reconstruction of statistically equivalent microstructures using only these abstractions in the absence of high-order information difficult.

Revived from near pseudoscience status during the "AI winter"[14], deep neural networks, which feature large model capacities and generalities, have stimulated a plethora of applications across different disciplines[15-21] (including materials science) in recent years. Existing deep learning-based approaches in materials science fall into two categories: material-system-dependent or -independent. Material-system-dependent approaches train deep learning models based on collected materials data, with their subsequent applications are often limited to the material system used for training. For instance, Cang et al.[22] extracted microstructure representations for alloy systems using convolutional deep belief networks. Their model[22] was trained with 100 images of size 200 x 200 pixels. While their model generated satisfactory reconstruction results for the chosen alloy material, their model was highly constrained to the type of alloy system used for the training set. Li et al.[23] developed a Generative Adversarial Network (GAN)-based model to learn the latent variables of a given set of synthetic microstructures, but their model needs to be retrained for application to a set of microstructures with significantly different dispersion. In contrast to these material-system-dependent approaches, *transfer learning* provides an alternative to capture microstructure characteristics without the need for training with a set of materials data (i.e., it is material-system-independent). Transfer learning[24-26] refers to the strategy of migrating knowledge for a new task from a related task that has already been learned[27]. In the context of microstructure analysis, deep-learning models trained for benchmark tasks using computer vision are fully or partially adopted to quantify microstructures or to address other complex challenges. For instance, DeCost et al.[28] utilized a transferred deep convolutional network to capture hierarchical representations of microstructures and then used these representations to infer the underlying annealing (i.e., processing) conditions. Lubbers et al.[29] adopted the VGG-19 model[30], trained on ImageNet[31], and used the activations of its network layers as microstructure representations to identify physically meaningful descriptors (e.g., orientation angles) via manifold learning. Nevertheless, none of these newly developed transfer learning-based approaches has addressed the challenge of microstructure reconstruction, where the extracted features from a network need to be reproduced in a statistically equivalent way. It should be noted that in Lubbers et al.[29], a prior texture representation based on the activations of deep convolutional layers previously developed by Gatys et al.[32] was implemented to synthesize visually similar microstructures with the same texture representation. However, the more challenging problem of achieving statistical equivalency of microstructures was not addressed in their work.

In the present study, a generalized transfer learning-based, training-free approach is proposed for reconstructing statistically equivalent microstructures from arbitrary material systems using a single given target microstructure. The input microstructure with labeled material phases is first passed through an encoding process to obtain a 3-channel representation in which material phases are distantly separated. In the meantime, the initial 3-channel representation of the reconstructed microstructure is randomly generated as the initialization. In each iteration of the reconstruction process, both of the 3-channel representations of the original and reconstructed microstructures are fed into a pre-trained deep convolutional network, VGG-19[30], and a loss function is utilized to measure the statistical difference between the original and the reconstructed microstructures. The gradient of the loss function with respect to each pixel of the reconstructed microstructure is computed via back-propagation and is then utilized in gradient-based optimization to update the reconstructed microstructure. Finally, the updated 3-channel representation of the reconstructed microstructure is propagated through a decoding stage via unsupervised learning to obtain the reconstructed microstructure with labeled material phases. In addition to visual similarity, statistical equivalence of the reconstructed microstructure is achieved by the encoding-decoding pair, which ensures sharp phase boundaries with correct labeling for the phase of each pixel. In addition, to ensure the computational viability of the proposed approach, model pruning is conducted on the transferred deep convolutional network. For validation, microstructures generated by differently pruned models are evaluated via visual inspection, numerical validation, and the calculation of receptive fields, which are defined as the regions in the input space that influence a particular convolutional neural network feature. The correlation between network layers and microstructure dispersion is also concurrently analyzed. Finally, as an extension, the knowledge learned in model pruning is utilized in determining the architecture and initialization conditions in developing a structure-property predictive model. A numerical validation using a small dataset of microstructures and their optical properties is conducted in order to verify the proposed structure-property modeling approach.

**Microstructure Reconstruction**

The proposed transfer learning-based approach for microstructure reconstruction migrates a pre-trained deep convolutional network model[30] created using ImageNet[31] – an auxiliary dataset which contains millions of regular images – and adds encoding-decoding stages before and after the deep convolutional model, as illustrated in Fig. 1.

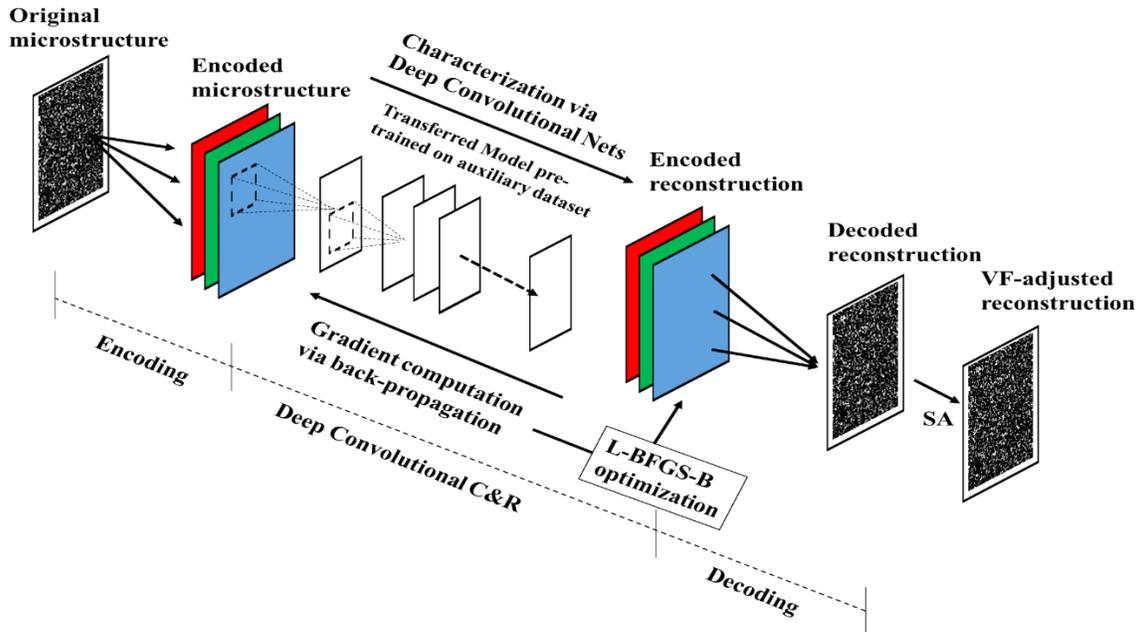

**Fig 1.** The work flow of the proposed approach for microstructure reconstruction.

**1) Encoding**: The transferred deep neural network has very strict requirements for data entry in terms of the image size and 3-channel representation alignment. Therefore, we encode the original microstructure in which each pixel is labeled with material phases into 3-channel representations so that the dimensionality of the input image fits the requirements of the transferred deep convolutional model. For ease of distinguishing individual phases after reconstruction, we employ maximize-minimum (maximin)[33] distance mapping from phase labels to the 3-channel representations.

**2) Gradient-based microstructure reconstruction:** Three steps are applied in transferring the deep convolutional neural network into microstructure reconstruction. a) **Removal of highest network layers**: It is well recognized that higher-level layers, particularly the last fully connected layer, are discriminators tuned specifically for the image classification task. For our new task of microstructure reconstruction using the transferred VGG-19 model based on non-material images, we eliminate the highest 7 layers (3 fully-connected and 4 convolutional layers with the associated pooling and dropout layers) (see details in the "Model Pruning" section below). b) **Gram-matrix computation**: Gram-matrix[32], which is usually used for measuring the differences in textures between images, is taken as the measurement of statistical equivalence between the original microstructure and the reconstruction. We implement its forward and backward computations (i.e. the calculation of Gram-matrix and its gradient) by customizing a computation unit and integrating it with the transferred model. On each layer of the convolutional deep network, first the Gram-matrices of the original and reconstructed microstructures are computed based on the activation values, then the differences of the Gram-matrices on the corresponding layers between the original and reconstructed microstructures are added as the optimization objective. c) **Gradient computation via back-propagation**: The state-of-the-art deep learning platform provides a fast and handy way of gradient calculation via back-propagation through the computation graph. The gradient of the objective (Gram-matrix difference) with respect to reconstruction image pixels is thus calculated via back-propagation. The gradient is then fed into nonlinear optimization (either L-BFGS-B[34] or Adam[35]) to update the reconstruction iteratively until convergence is found. It is noted that stochasticity of the microstructure reconstructions is achieved by random initialization of the microstructure image before the back-propagation operation.

**3) Decoding:** after obtaining the 3-channel representation of the reconstruction, an unsupervised learning approach is used to convert the 3-channel representation back to the desired representation: images with labeled material phases. Furthermore, considering that volume fraction (VF) is a critically important microstructural descriptor, a Simulated Annealing-based VF matching process is exercised at the end to ensure the reconstructed image has the same VF as the original through erosion or dilation.

**Model pruning of the transferred deep convolutional network**

While the proposed microstructure reconstruction approach is capable of generating not only visually similar but also statistically equivalent microstructures, its consumption of computational resources is significant, which hinders its wide application on computational platforms with limited capacity. Two major bottlenecks are the GPU memory consumption and the number of back-propagation operations. High GPU memory consumption would result in numerical errors in lower-end computational platforms, which hinders the wide application of the proposed approach, and a great number of back-propagation operations would significantly slow down the speed of gradient computations. Given that both GPU memory consumption and the number of back-propagation operations are affected by the depth of a deep convolutional network, we set our objective to reduce the hierarchical depth of the transferred model for computational economy and efficiency. The model pruning is implemented for this purpose in two steps: 1) we gradually remove the top layer(s) from the existing model and generate reconstructed microstructures, and 2) we analyze the trade-off between model depth and reconstruction accuracy by both visual inspection and numerical validation using two-point correlation function and lineal-path correlation function. We also compute the receptive field of each pruned model and investigate how dispersion in the microstructure determines the size selection of receptive fields, which plays a decisive role in model pruning.

**Structure-property prediction**

While microstructure reconstruction approaches are capable of generating statistically equivalent realizations, it is always computationally costly to simulate the material properties of these microstructures via Finite Element Analysis (FEA). Fortunately, as the data of structure-property mapping is accumulated, it becomes feasible to train machine-learning models, which have significantly shorter runtime than FEA, to replace the physical simulations. As deep convolutional networks become increasingly prevalent, a common practice of building a structure-property predictive model is to transfer an existing pre-trained model either fully or partially, and fine-tune the weighting using back-propagation[36]. A crucial choice in this transfer learning process is to determine which part of the pre-trained network should be adopted. In the existing studies that transfer the pre-trained model, this choice varies a lot. For instance, Yosinski et. al[36] adopted the full AlexNet[18] in an image classification task while Li et al.[23] only used the first four convolutional layers in a pre-trained Generative Adversarial Network (GAN) model to build structure-property prediction of optical materials.

While the determination of which portion of the pre-trained model to adopt is usually subjective, the aforementioned model pruning study provides an objective guideline. Since the pruning of the reconstruction model reveals the necessary part of the pre-trained model to be transferred, this results in a rule specifying the network architecture and initialization conditions. In this paper, the proposed approach is compared with two different network architectures, demonstrating enhanced stability for the former.

# Results and Discussion

## Material systems

Two different datasets have been prepared for the tasks of microstructure characterization and reconstruction (MCR) and structure-prediction, respectively. First, a dataset of microstructure images obtained by state-of-the-art microstructure imaging techniques, covering carbonate, polymer composites, sandstone, ceramics, a block copolymer, a metallic alloy and 3-phase rubber composites (the 1st row in **Fig. 2**) has been collected for demonstration and validation of the proposed MCR approach. Given the great variety of microstructural morphologies, this dataset provides a comprehensive test-base for comparing our proposed approach to other MCR approaches. Among all test samples, special attention is given to two challenging systems – 2-phase block copolymer and 3-phase rubber composites. The block copolymer sample has a fingerprint-shaped microstructure, in which anisotropy is observed locally whereas isotropy holds globally. In contrast, the rubber composite sample has higher local isotropy, yet its three-phase nature is difficult to capture using any prior approach. For the second task of structure-property prediction, an additional dataset consisting of structure-property pairs was obtained by generating 5,000 microstructure patterns using the Gaussian Random Field [9](GRF) method (a popular and computationally efficient choice to generate microstructures of optical materials) with a wide range of correlation parameters, followed by subsequent simulation of their light absorption rates at a wavelength of 600nm using Rigorous Coupled Wave Analysis (RCWA). RCWA is a Fourier space-based algorithm that provides the exact solution to Maxwell's equations for electromagnetic diffraction. While we set the diffraction order in RCWA such that each simulation takes less than 5 minutes to complete, more accurate simulation solutions can be obtained by choosing higher diffraction orders.

## Validation of Microstructure Reconstruction Results

The quality of microstructure reconstruction is assessed both quantitatively based on numerical metrics and qualitatively through visual inspection. As the two-point correlation function is the most commonly used statistical function to evaluate microstructure reconstructions[8,10], we adopt it as one of the quantitative evaluation metrics in the present work. However, per Torquoto[37], the two-point correlation function itself is not sufficient in evaluating the statistical equivalence of microstructures. In this work, the lineal-path correlation function[38] is used as an additional metric to quantify the statistical similarity between the original and reconstructed microstructures[10]. Since most statistical functions are reduced representations of microstructures, they cannot reveal the microstructure characteristics completely. To this end, visual inspection was also conducted as a complementary validation to the numerical comparisons.

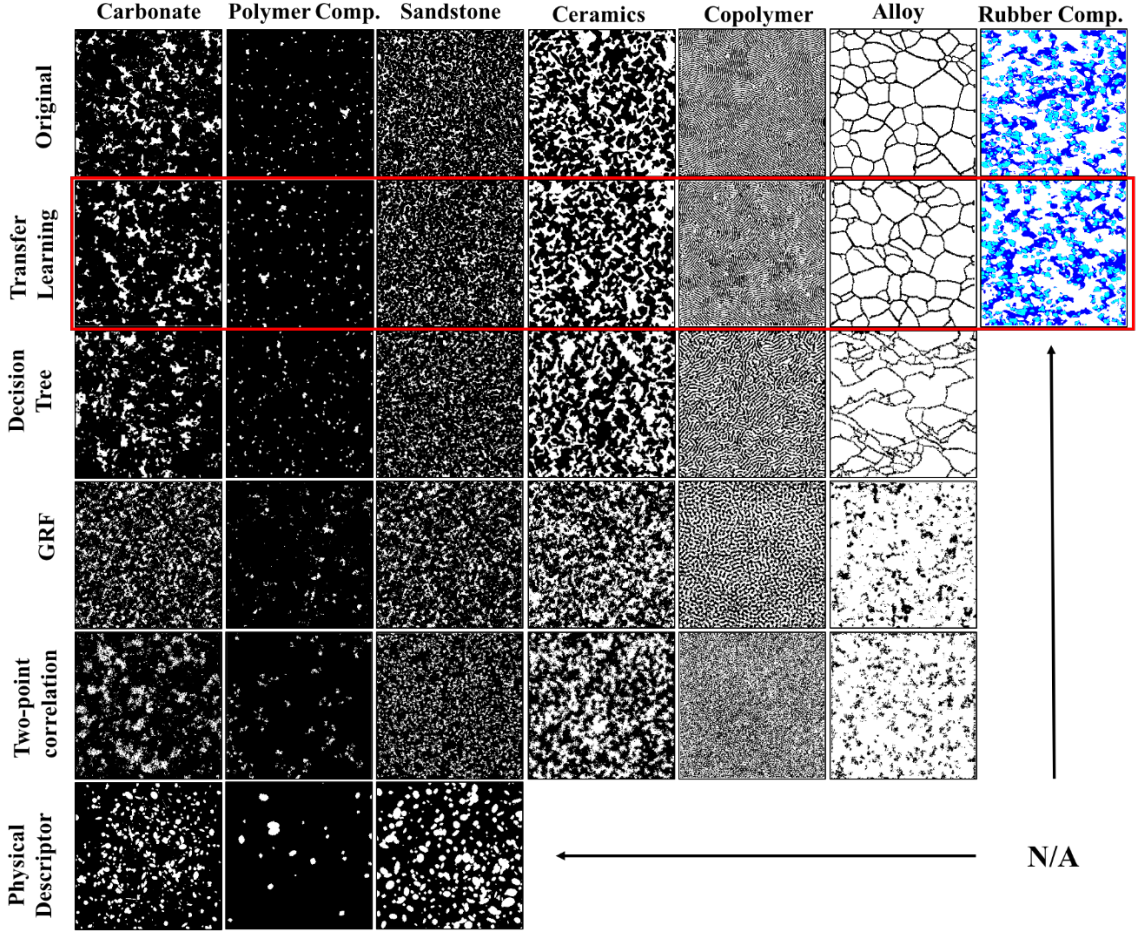

**Fig. 2** The comparison of the original microstructures and their corresponding reconstructions using different approaches. The proposed transfer learning approach reconstructions are presented in the second row, highlighted in red. N/As represent the cases where the microstructure in that column cannot be reconstructed by the approach specified for that row.

As depicted in **Fig. 2**, in addition to the proposed transfer learning-based approach (Row 2), four existing MCR approaches (i.e., decision tree-based synthesis[10], Gaussian Random Field[9,39,40], two point correlation[7,41], and physical descriptor[8]) are applied to each of the microstructures in the collected data, except for the three-phase rubber composite sample in the last column since none of the commonly used approaches can process the multi-phase microstructure of the rubber composite (three materials phases). The discrepancy between the correlation functions of the original microstructure and those from reconstructions is measured by

$$\varepsilon = \frac{s_1}{s_2} \times 100\%, \tag{1}$$

where $s_1$ is the area between the two correlation functions and $s_2$ is the area under the correlation function of the original microstructure. The error rates of the reconstruction for each method in each material sample are illustrated in **Tables 1** and **2**. It should be noted that in the copolymer and ceramic samples, the white material phase is almost all connected, and thus it is inappropriate to apply the physical-descriptor based approach. In addition, for the alloy material system (Fig. 2, column 6), the proposed approach is significantly better than other microstructure reconstruction approaches in generating visually satisfactory reconstructions. Since visual similarity between the original and reconstructed microstructures is a necessary qualitative criterion to validate the equivalence of microstructures, we do not conduct further numerical validation for the alloy material system in the later part of this paper.

From **Table 1**, we find that the proposed transfer learning based approach for using convolutional deep networks outperforms all other reconstruction approaches in four out of the five material systems being numerically evaluated. For the sandstone sample, the accuracy of reconstruction using the proposed approach is just slightly lower than that of the two-point correlation function based approach. One may expect that the error evaluated using the two-point correlation function should be the smallest when the two-point correlation approach is used for reconstruction because the metric is directly used as an objective. However, the two-point correlation function based reconstruction uses simulated annealing, which yields difficulty in converging to the global minimum, leading to poorer performance. Moreover, while the reconstructions from the GRF and two-point correlation function based approaches on the copolymer material system achieve a relatively low error rate (0.85% and 1.20%, respectively), those reconstructions are visually different from the original microstructure. This again verifies Torquato's proposition[37] that two-point correlation only partially reveals the statistical equivalence of the original microstructure and its reconstructions. Finally, the error rates for the polymer composite material system are observed to be higher than those of other systems. For the low-loading polymer nanocomposite material system, the values of $s_2$ in **Eq. 1** are lower than those for the other material systems studied in this work. Therefore, a slight difference between the correlation functions ($s_1$) would lead to a significantly larger error rate.

**Table 1**. Error rate (%) of two-point correlation function for reconstructions using different approaches for various material systems (bold font indicates the method with the lowest error rate for each material system). The method presented in this work is highlighted in red.

| Method \ Material | carbonate | ceramics | sandstone | copolymer | Polymer comp. |
|---|---|---|---|---|---|
| Transfer Learning | **3.91** | **1.00** | 1.74 | **0.78** | **6.0** |
| Decision tree | 4.76 | 1.09 | 8.51 | 1.62 | 13.71 |
| GRF | 9.62 | 2.06 | 2.58 | 0.85 | 30.53 |
| Two-point correlation | 5.16 | 1.13 | **1.17** | 1.20 | 13.2 |
| Physical descriptor | 6.92 | N/A | 6.47 | N/A | 12.51 |

**Table 2** illustrates the error rate evaluated using the lineal-path correlation function. Three major findings are summarized from this comparison. 1) The transfer learning-based approach achieves a low error rate (<8%) in all five samples, while the performance of the other four methods varies significantly across different material systems. 2) While the two-point correlation function based approach reaches very low error rates in Table 1, its error rates of the lineal-path function is very large. This result is reasonable since the two-point correlation function based approach applies pixel-switching in its reconstruction process, but the connectivity in the clusters is not guaranteed. 3) While the transfer learning-based approach demonstrates superiority in terms of generality, the decision-tree based approach is a very competitive also achieves very low error rates in three of the five samples.

**Table 2**. Error rate (%) of linear-path correlation function for reconstructions using different approaches for distinct material systems (bold fond indicates the lowest error rate, red highlight indicates the proposed method.)

| Material Method | carbonate | ceramics | sandstone | copolymer | Polymer comp. |
|---|---|---|---|---|---|
| Transfer learning | 7.63 | **1.31** | 3.61 | **6.38** | **3.58** |
| Decision tree | **3.26** | 1.69 | **3.10** | 50.30 | 14.71 |
| GRF | 59.38 | 49.28 | 47.79 | 27.41 | 59.58 |
| Two-point correlation | 45.59 | 37.92 | 24.19 | 8.09 | 30.65 |
| Physical descriptor | 18.08 | N/A | 28.11 | N/A | 12.86 |

While the error rates of the proposed transfer learning-based convolutional network approach and the two-point correlation based approach are very close in a few cases (e.g. copolymer), their reconstructions could significantly differ from visual inspection. This again implies that while lower-order statistical functions can capture lower-order statistical equivalence, high-order metrics are needed to completely assess the statistical equivalence.

Since both the two-point correlation function and lineal-path correlation function are low-order representations of microstructures, they do not fully capture the high-order characteristics of the original and reconstructed microstructures. To this end, we also visually inspected the reconstructions of different material systems (**Fig. 2**) and compared our findings to the results in **Tables 1 & 2**. In general, the visual similarity between the original microstructures and the reconstructed ones agrees with the error rates in **Tables 1 & 2**, with the exception of the block copolymer reconstruction using the two-point correlation function based approach. In this case, the reconstruction achieves 1.20% error rate in the two-point correlation function and 8.09% error rate in the lineal path correlation function. However, the reconstruction looks like a random white noise image by visual inspection. This finding again confirms Torquato's proposition[37] that low-order statistical functions are not capable of representing the microstructure completely.

In addition to demonstrating the advantages through quantitative comparative studies (**Fig 2 & Tables 1-2**), we demonstrate the versatility of the proposed approach through analyzing complex microstructures, such as those in block copolymer and 3-phase rubber composite samples. As illustrated in **Fig 3**, the original fingerprint-shaped copolymer sample has very different local anisotropy at different locations, whereas the global isotropy holds. The proposed deep convolutional network-based approach can accurately reproduce this characteristic in its reconstruction, while the decision tree-based approach generates a diagonally oriented anisotropic reconstruction.

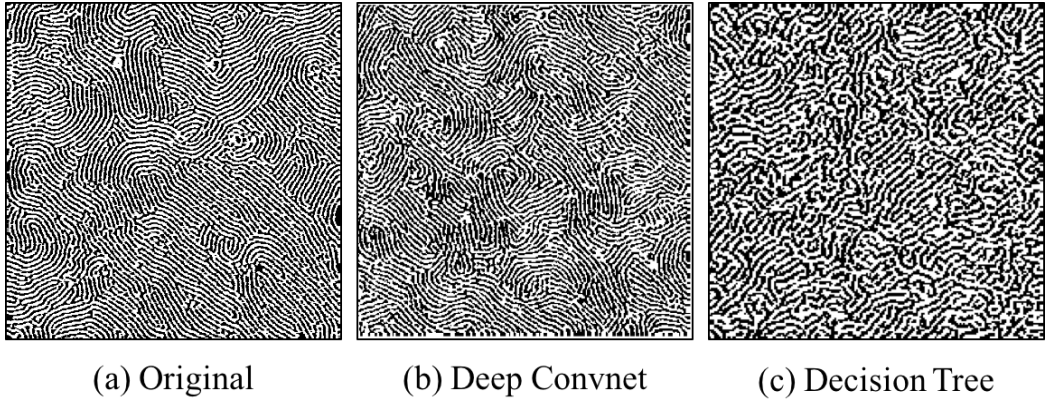

| (a) Original | (b) Deep Convnet | (c) Decision Tree |

**Fig. 3**. Original microstructure of block copolymer sample and its reconstructions using the proposed deep convolutional network-based approach and decision tree based approach.

Advantages of the proposed approach are also demonstrated in **Fig. 4** by analysis of a rubber composite sample that consists of two rubber phases (Butadiene rubber (BR, white) and Styrene-Butadiene rubber (SBR, blue)) with one filler phase (carbon black (CB, cyan)), at two different carbon black compositions. Given the multiphase nature of this material, the statistical equivalence of the original microstructures (**Figs 4(a) & 4(c)**) and their reconstructions (**Figs 4(b) & 4(d)**) is evaluated using the two-point correlation function in the one-vs-rest manner: specifically, the three-phase microstructure is first binarized into three binary images (BR vs. the rest, SBR vs. the rest and CB vs. the rest). Then the correlation function is applied to the binary images in order to validate the statistical equivalence. Using this method, the statistical equivalence of the original microstructures and their reconstructions are validated (**Table 3**).

**Table 3**. Error rate (%) of the two-point correlation function for reconstructions on three-phase rubber composite samples in **Fig 4** using the proposed deep convolutional network-based approach.

| Image | Composite #1 (Fig 4(a) & (b)) | Composite #2 (Fig 4(c) & (d)) |
|---|---|---|
| BR-vs-rest | 0.77 | 0.80 |
| SBR-vs-rest | 2.25 | 1.99 |
| CB-vs-rest | 3.67 | 5.25 |

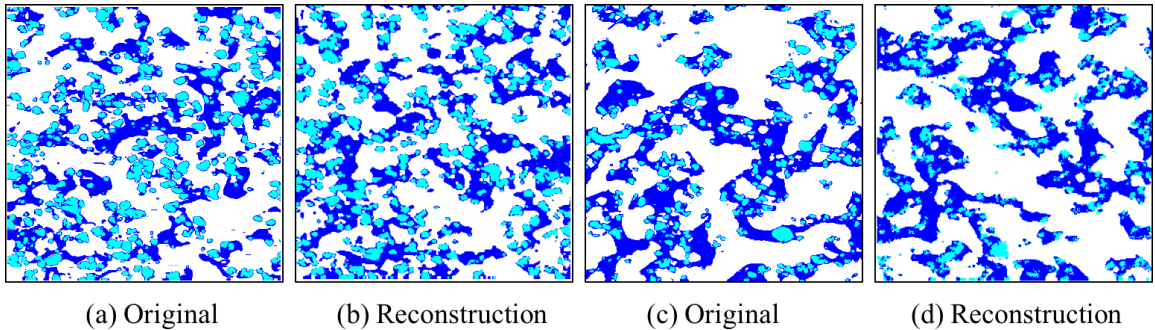

| (a) Original | (b) Reconstruction | (c) Original | (d) Reconstruction |

**Fig 4**. Original microstructures of 3-phase rubber composite and their reconstructions using the proposed deep convolutional network-based approach. (a&b) Original and reconstructed microstructures of BR(35.7 wt%)/SBR(35.7 wt%)/CB(28.6 wt%) sample. (c&d) Original and reconstructed microstructures of BR(41.7 wt%)/SBR(41.7 wt%)/CB(16.6 wt%) sample. Color map: Butadiene rubber (BR, white color), Styrene-Butadiene rubber (SBR, blue color)), carbon black (CB, cyan color).

## Numerical pruning and understanding the network model hierarchy

While it is shown that the proposed approach is capable of reconstructing statistically equivalent microstructures accurately for a wide range of material systems, the application of the proposed approach is potentially limited because of its high computational cost (primarily GPU memory consumption and the number of operations in back-propagation.) In our experiments, the loading of full VGG-19 model consumes 11541MB GPU memory on a Nvidia GeForce Titan Xp graphic card and leads to a significant amount of back-propagation operations. Therefore, in this section, model reduction is studied by eliminating some network layers to increase computation efficiency and viability. Noticing that, different computational platform may have very different computational performance for the same model, in this study, the number of weight parameters is used to measure the model complexity.

The model pruning in this work is achieved by first gradually eliminating high-level layers, followed subsequently by elimination of low-level layers from the transferred deep convolutional network. This sequence in layer removal not only keeps the lower part of the network architecture intact, but also aligns with the understanding of network architectures presented by Yosinski et al.[36] In other words, higher-level layers are likely to be high-level concept discriminators for specific tasks. Thus, their elimination may impact the reconstruction less significantly than the removal of lower-level layers (i.e., the ones close to the image), which are usually interpreted as general local feature extractors similar to Gabor filters[42] or color blobs. In the vanilla version of the proposed approach, the first convolutional layer and the first four pooling layers are included in the loss function. The inclusion of these five layers essentially requires the loading of the first 12 convolutional layers, which introduce 10,581,696 parameters (not counting the biases). The removal of the highest pooling layer (pooling_4 in VGG-19) of the five layers reduces the number of included convolutional layers to 8, which have 2,324,160 parameters (21.96% of the previous one). Further elimination of the other pooling layers (pooling_3, pooling_2 and pooling_1) would reduce the number of convolutional layers to 4, 2 and 1 respectively, which corresponds to 259,776/ 38,592/ 1,728 weight parameters (2.45%, 0.36% and 0.02% of the first one), respectively. **Fig. 5** illustrates the reconstructions using different selections of layers. From this comparison, we find three important results. 1) The elimination of the highest pooling layer has an insignificant impact on the reconstruction results. This result is expected, as those higher pooling layers are discriminators specifically tuned for the original AI task (i.e. image classification for ImageNet dataset). 2) From the comparison (**Fig 5**. A-3 & A-4, B-3 & B-4), the removal of the third pooling layer results in the loss of long-distance dispersion equivalence. Specifically, in **Fig. 5** (A-4) global variation of local anisotropies is lost, while variation of cluster-cluster distances is decreased in **Fig. 5** (B-4). 3). From the comparison (**Fig 5**. A-4 & A-5, B-4 & B-5), the elimination of the lowest pooling layer leads to significant loss of short-distance (local morphological) equivalence. Our observations further validates our hypothesis that in transferring deep learning models, the highest neural network layers (i.e. layers higher than pooling_3) may be eliminated because they are discriminators for the original ImageNet[43] image classification task but are not useful for microstructure reconstruction. In contrast, network layers lower than pooling_3 need to be retained to keep dispersive characteristics in the reconstructed microstructure. **Fig. 5** (C&D) illustrates the reconstruction error rates (Eqn. 1) computed using two-point correlation function and lineal-path correlation function. It is observed that removal of pooling_3 and pooling_2 would not affect the reconstruction accuracy significantly. Noticing that neither of the two correlation functions could fully capture microstructure characteristics, in determining the optimal pruned network architecture, we still retain the layers that are necessary for both visual similarity and statistical equivalence (i.e. layers lower than pooling_3).

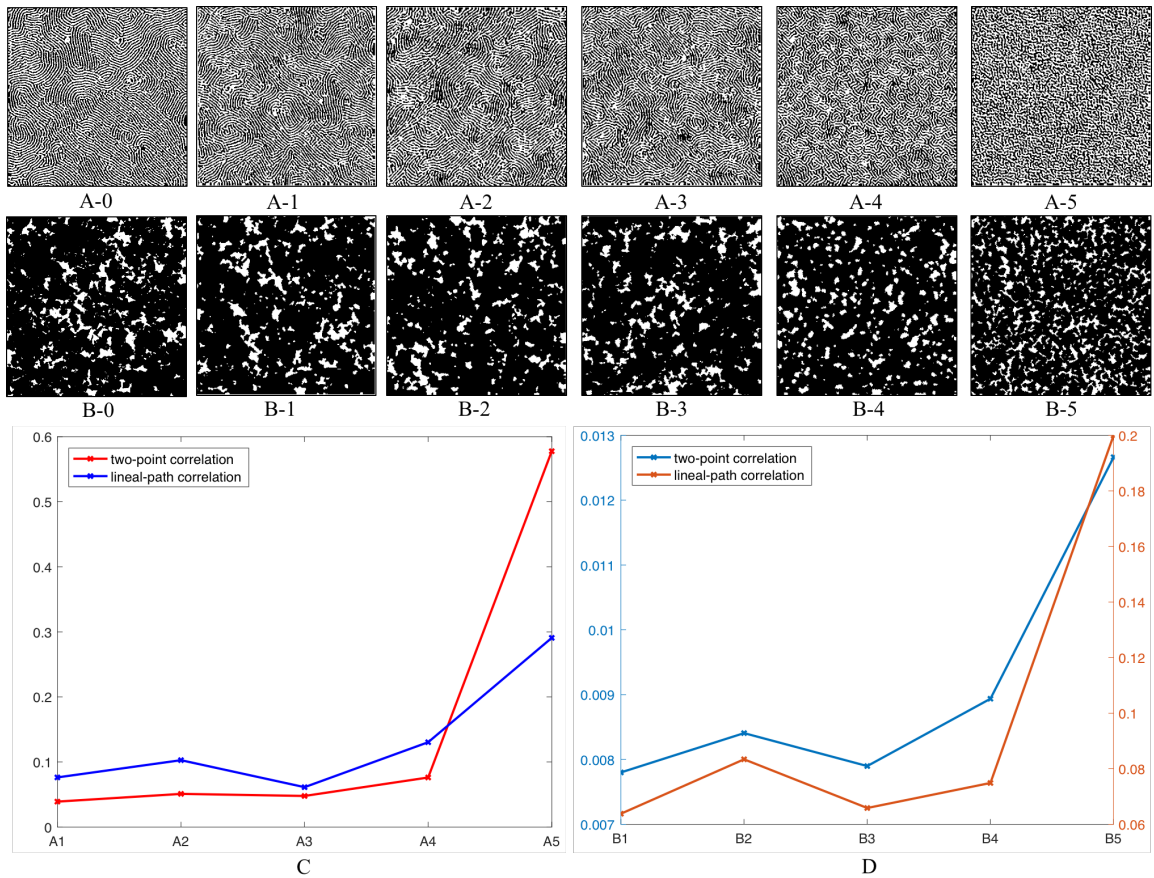

**Fig. 5**. Microstructure reconstructions for copolymer and carbonate using different selections of neural network layers in Gram-matrix matching. (Figure index: A ~ copolymer, B ~ carbonate, 0 ~ original microstructure, 1 ~ 4 lowest pooling layers + lowest convolutional layer, 2 ~ three lowest pooling layers + lowest convolutional layer, 3 ~ two lowest pooling layers + lowest convolutional layer, 4 ~ the lowest pooling layer + lowest convolutional layer, 5 ~ lowest convolutional layer only).C. Comparison of the reconstruction errors of each pruned model for copolymer sample using correlation functions. D. Comparison of the reconstruction errors of each pruned model for carbonate sample using correlation functions.

In addition to the numerical study illustrated above, the model pruning is also analyzed from the perspective of receptive fields. A receptive field is a significant concept in deep convolutional networks and is defined as the region in the input space that influences a particular convolutional neural network feature. As all the convolutional filters in the VGG-19 model are 3x3 pixels, it is relatively straightforward to compute the receptive fields for each layer (**Table 4**). The sizes of the receptive fields could be interpreted as follows: for the lowest convolutional layer (conv_1-1), varying each entry of its output can affect a small region of 3x3 pixels, while altering the output of the pooling_4 layer leads to the influence of a large area of 160x160 pixels (the full microstructures in this work are 256x256 pixels). The sizes of the receptive fields for each layer also reveal their individual roles in controlling the microstructure reconstruction. Specifically, higher-level layers (e.g. pooling_3) control the long-distance dispersion in the microstructure, while lower-level layers (e.g. conv_1-1 and pooling_1) specify local geometries. This again validates our findings in the comparison of reconstructions in **Fig. 5**. For the two material systems in **Fig. 5**, a 72x72 pixel window from the microstructure is capable of capturing most of the statistical characteristics; therefore, layers higher than pooling_3 could be eliminated while retaining the quality of the microstructure reconstruction.

Table 4. Receptive field for each layer used for computing loss function in the proposed approach

| Layer | Conv_1-1 | Pooling_1 | Pooling_2 | Pooling_3 | Pooling_4 |
|---|---|---|---|---|---|
| **Receptive field** | 3x3 | 10x10 | 28x28 | 72x72 | 160x160 |

**Structure-property Prediction**

In addition to using deep learning for microstructure reconstruction, a natural extension is to employ the architecture of deep convolutional networks for analyzing structure-property relationships of advanced materials. It has been found by Yosinski et al.[36] and LeCun et al.[21] that transfer learning (i.e. using pre-trained weights to initialize the network) improves the stability and accuracy of the predictive model. Despite these successes, there is no rule to determine how many layers to transfer, thereby introducing subjective choice. Typically, the inclusion of more pre-trained layers increases the flexibility of the network model but increases the associated computational cost and the likelihood of over-fitting. To resolve this issue, the model pruning investigated in the previous section is used to identify the necessary pre-trained layers for describing dispersive characteristics in microstructures and thus provides a guideline for this determination. Specifically, we propose the general rule of determining the number of transferred layers as follows: the remaining layers in the pruned microstructure reconstruction model are regarded as necessary ones to describe microstructure characteristics; therefore, they also need to be adopted in developing structure-property predictive models. In the context of VGG-19 models, all the layers beyond pooling_3 are discarded in pruning and the remaining 8 convolutional layers and 3 pooling layers are utilized to initialize the neural network for structure-property predictions.

To demonstrate the effectiveness of this approach, a numerical study is conducted to develop a structure-property predictive model for optical microstructural materials. 250 microstructures of size 128x128 pixels are generated using the Gaussian Random Field approach[9], and their corresponding optical absorption properties (a scalar value between 0 and 1) are simulated using Rigorous Couple Wave Analysis[44] (RCWA). This dataset is split into 200 and 50 microstructures for training and testing, respectively. **Fig. 6** shows several examples of the generated microstructures. The architecture of the neural network is constructed using the layers lower than the pooling_3 layer in VGG-19. The output of the pooling_3 layer is flattened, followed by two fully connected layers of size 2048 pixels and 1024 pixels with Rectified Linear Unit (ReLU) and dropout (p=0.5) operations. The weights in the transferred layers are initialized using the pre-trained weights in the VGG-19 model while the remaining ones are initialized randomly. Two additional experiments are also conducted as control groups: **group 1** – instead of adopting layers lower than pooling_3, we only transfer the ones below pooling_2, and the rest of the settings are kept the same as the proposed approach; **group 2** – layers lower than pooling_4 are adopted, and the same settings for fully connected layers are used. Adam optimizer (learning rate = 0.0005, beta1=0.5, beta2=0.99) is used to fine-tune the parameters. The mini-batch size is set as 50 and the number of epochs is 4,000. For each group, the training is repeated 15 times to investigate accuracy and stability.

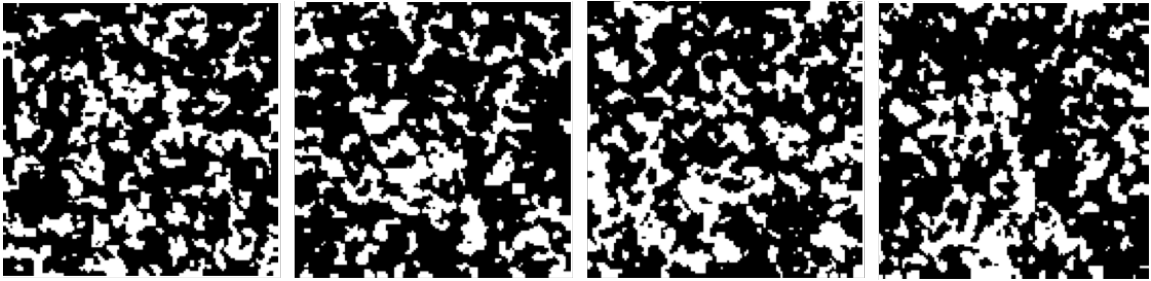

**Fig. 6**: Examples of the generated microstructures for developing structure-property predictive model.

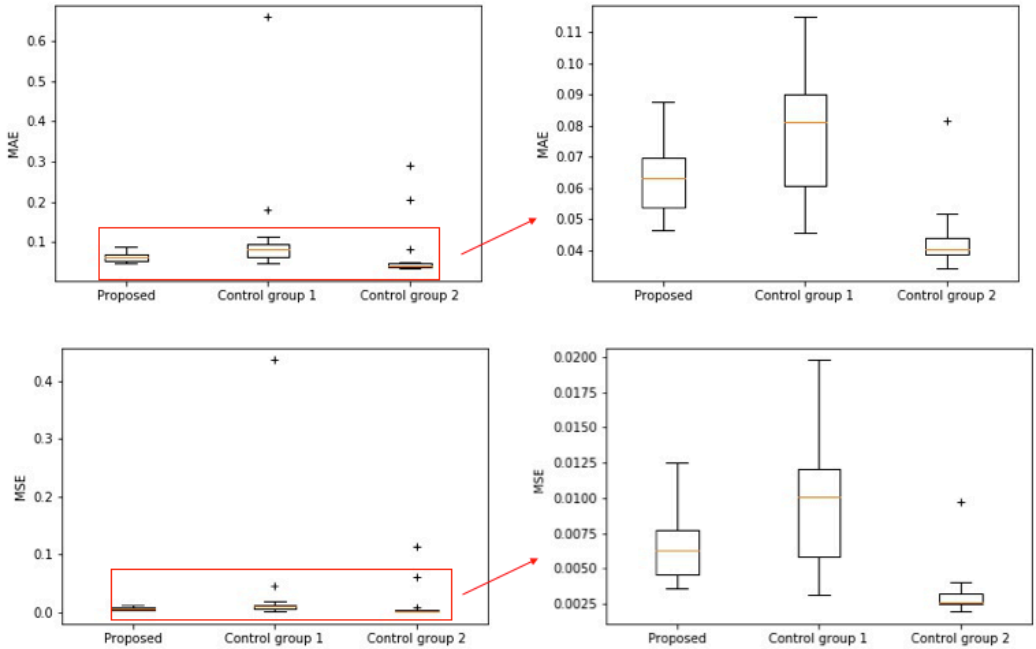

**Fig. 7** The comparison of the mean-squared-error (MSE) and the mean-absolute-error (MAE) between the proposed approach and two control groups.

**Fig. 7** shows the results of the proposed approach and two control groups. Comparing to the proposed approach, the control group 1 is observed to be under-fitted (i.e., higher mean error and larger variance of the error), while the control group 2 shows a higher likelihood of overfitting (i.e., more outliers with large associated error). This comparison validates the significance of the knowledge learned from microstructure reconstruction model pruning, and it also validates the proposed structure-property modeling approach.

## Conclusion

Microstructure reconstruction and structure-property prediction are two challenging but advantageous tasks in computational materials science. In this work, a transfer learning based approach for reconstructing microstructures is first proposed. A comprehensive comparison of results for multiple materials between the proposed approach and existing approaches is conducted to demonstrate the accuracy and generality of the proposed approach. To reduce the computational cost, the transferred deep convolutional network is pruned, and the understanding of the correspondence between neural network layers and long/short-range dispersions in microstructures are drawn by visual inspection and analyzing the receptive fields. The knowledge learned in model pruning also guides the determination of the pre-trained layers to be transferred in developing structure-property predictive models. In summary, the proposed approach provides an end-to-end – i.e. image-to-image for reconstruction or image-to-property for property prediction – and off-the-shelf solution which generalizes well and requires minimum prior knowledge of material systems.

Despite the advantages demonstrated in this work, the present approach has some potential limitations. For instance, while the Gram-matrix matching ensures the statistical equivalence in stochastic microstructure reconstructions, it is not guaranteed to be applicable in deterministic microstructures such as periodic crystallographic structures. The reconstruction of these deterministic microstructures may be handled by adding customized loss function terms to the proposed approach. In addition, the adoption of pre-trained ImageNet deep convolutional network implicitly constraints the application of the proposed approach on 2D microstructures. 3D microstructure reconstruction tasks may be solved by extending the proposed transfer learning strategy using existing 3D convolutional network models.[45] From the perspective of deep learning, advanced deep learning models such as ResNet[46] may further improve the results.

## Method
### Encoding via Maximizing Minimum Distance

Microstructures are typically represented as NxM matrices (where N and M correspond to the height and width of the microstructure image, respectively). The first step is to convert the NxM matrices to the 3-channel representations that can be fed into the transferred deep convolutional model. While there are a variety of mapping methods for this conversion, here we suggest an encoding strategy that maximizes the minimum Euclidian distance between the encoded phase coordinates. This encoding strategy is chosen for the ease of distinguishing individual phases after the gradient-based optimization for reconstruction in the encoded space is carried out.

Maximization of the minimum distances between a number of points in the feature space has been solved typically by gradient-based search algorithms or stochastic search algorithms such as simulated annealing[47]. It can also be formulated as an NP-complete, independence in geometric intersection graphs problem, which can be addressed by approximate algorithms[33]. However, given that in most cases, the number of different material phases in an original microstructure is not large, it is not necessary to pursue the farthest separations as long as the phase clusters after reconstruction can be properly distinguished. Therefore, for material systems which have no more than three (3) material phases such as in this work, we take a simpler approach – Latin Hyper-cube Sampling[48] (LHS). Specifically, by setting the number of sampling points to be equal to that of distinct material phases in the original microstructure, LHS samples with maximin distance criteria would create a 3-vector representation for each material phase. Then, for each pixel in the NxM microstructure, we replace the original scalar phase label by the 3-vector, which leads to NxMx3 matrix representations.

### Deep Convolutional Characterization and Reconstruction

While a lot of models have been developed for the ImageNet task such as GoogleNet[49] and ResNet[46], the VGG-19 model[30] is selected in this work because of its structural simplicity and regularity. The original VGG-19[30] model has 19 layers (3 fully connected layers and 16 convolutional layers). In transferring this model, all layers beyond the 2$^{nd}$ highest pooling layer are first eliminated (i.e. 1 fully-connected layer, 1 pooling layer and 3 convolutional layers). Both the network structure and the network parameters from the VGG-19 model are inherited as the transferred deep convolutional model in this work.

Microstructure reconstruction using the transferred deep convolutional model is essentially a gradient-based optimization process. The objective function to be minimized is the sum of Gram-matrix differences on the selected neural network layers, and the variables to be optimized are the pixel values in the microstructures. The optimization process can be decomposed into three steps: 1) Initialization: an NxMx3 matrix is initialized randomly with uniform distribution for each entry in the microstructure. Different initializations will result in different statistically equivalent microstructure reconstructions. 2) Forward-propagation: At each iteration of optimization, the values in NxMx3 representations of the original and the reconstruction are forward-propagated simultaneously through the deep learning network, creating corresponding activation values on each layers. 3). Back-propagation: Gram-matrix[32] on selected layers are matched between the reconstruction and the original microstructure to find the difference (i.e., loss). The gradient of the loss with respect to each pixel in reconstruction is then computed via back-propagation using GPU, and it is then fed into a nonlinear optimization algorithm to update the pixel values of the microstructure reconstruction. Steps 2) and 3) are then executed iteratively until the solution converges to a local optimal state of the microstructure reconstruction.

The convolutional deep neural network in the present approach is composed of two sets of computation units: regular units (convolutional operation, Rectified Linear Unit transformation, and pooling operation) and customized units (Gram-matrix related computations). While the back-propagation of regular units are well integrated in the popular deep learning platforms, the Gram-matrix related derivations are still needed for the implementation in customized units. Here we demonstrate the derivation briefly. Let $\boldsymbol{x}$ and $\tilde{\boldsymbol{x}}$ denote the original and reconstructed microstructure in the encoded space at iteration #n, respectively. $\boldsymbol{x}$ and $\tilde{\boldsymbol{x}}$ are first passed through the transferred convolutional network for activating feature maps $F^i$ of layer $i$. Then in each layer $i$ of the network, $\boldsymbol{x}$ and $\tilde{\boldsymbol{x}}$ will activate a stack of feature maps $F^i, \tilde{F}^i \in \mathcal{R}^{N_i \times M_i}$, where $N_i$ is the number of filters and $M_i$ is the size of the vectorized feature maps in layer $i$. Let $F_{jk}^i, \tilde{F}_{jk}^i$ denote the activations of the $j^{th}$ filter at position k in layer $i$ for $\boldsymbol{x}$ and $\tilde{\boldsymbol{x}}$. The Gram matrix[32] of both microstructures is defined as the inner product between feature map $p$ and $q$ in layer $i$:

$$G_{pq}^i = \sum_r F_{pr}^i F_{qr}^i \qquad (2)$$

$$\tilde{G}_{pq}^i = \sum_r \tilde{F}_{pr}^i \tilde{F}_{qr}^i \qquad (3)$$

The contribution of the loss in layer $i$ is:

$$E_i = \frac{1}{4N_i^2 M_i^2} \sum_{j,k} (G_{jk}^i - \tilde{G}_{jk}^i)^2 \qquad (4)$$

The total loss is:

$$L = \sum_i \boldsymbol{E}_i, \qquad (5)$$

Next, a gradient-based optimization with with aim of minimizing the total loss between the original and the reconstructed microstructures is utilized in order to update the reconstructed microstructure. The gradient $\frac{\partial L}{\partial x}$ is decomposed by the chain rule as:

$$\frac{\partial L}{\partial x} = \frac{\partial L}{\partial E} \cdot \frac{\partial E}{\partial \tilde{F}} \cdot \frac{\partial \tilde{F}}{\partial x} \tag{6}$$

where

$$\frac{\partial L}{\partial E_i} = w_i \tag{7}$$

$$\frac{\partial E_i}{\partial \tilde{F}_{jk}^i} = \begin{cases} \frac{1}{N_i^2 M_i^2}\left[\left(\tilde{F}^i\right)^T (G^i - \tilde{G}^i)\right], & if \ \tilde{F}_{jk}^i > 0 \\ 0 & , \ \text{otherwise} \end{cases} \tag{8}$$

and $\frac{\partial \tilde{F}}{\partial x}$ is automatically handled by back-propagation in Caffe. The gradient $\frac{\partial L}{\partial x}$ is then fed into the Limited-memory Broyden-Fletcher-Glodfarb-Shanno algorithm[34] with bound constraints of [0, 255] (L-BFGS-B) or Adam optimizor[35] on each encoded dimension to minimize the total loss $L$. At the end, convergence at local minimum is achieved and each pixel will be assigned a 3-vector by L-BFGS-B. In other words, at the end of the optimization, a NxMx3 matrix will be obtained for the next step of decoding.

**Decoding Reconstruction via Unsupervised Learning and Simulated Annealing**

The reconstructed microstructure in the encoded space obtained from the previous step is essentially an NxMx3 matrix with each entry ranging from 0 to 255. To generate a microstructure image that is compatible to further numerical analysis such as Finite Element simulations, it is critical to convert the NxMx3 image back to the NxM image with each pixel appropriately labelled with material phases. In other words, for each location in the microstructure, its current representation of 3-vector needs to be replaced by a scalar label that indicates material phase. Since the reconstruction in the encoded space obtained by L-BFGS-B or Adam optimization is a local minimum, there is no guarantee that the 3-vectors of the reconstructed pixels are still exactly at the coordinates we sampled in the encoding step. Nevertheless, it is observed that 3-vectors of pixels for the same material phase are still clustered. Hence, we apply an unsupervised learning approach (K-means clustering) to separate the reconstructed pixels into K groups, where K is the number of material phases counted in the encoding process.

It should be noted that K-means clustering does not enforce the ratio of pixels' partition for each material phase, so it is possible that the volume fraction of each cluster is slightly different from that of the original microstructure. Considering that volume fraction (VF) is a key feature for material systems, such as polymer composites or carbonates, the last step of the algorithm is to compensate the discrepancy of VF between the original and the reconstructed microstructures by switching pixels' phase label on the boundary. Herein we utilize a stochastic optimization approach, simulated annealing (SA), to match the phase VFs with those in the original microstructure.

# Acknowledgement

This work is supported by Center for Hierarchical Materials Design (ChiMaD NIST 70NANB14H012), Design of Engineering Material Systems ( DEMS CMMI-1334929(NU), CMMI-1333977 (RPI)) and the Goodyear Tire and Rubber Company.

# Author contributions

W. Chen, X. Li, Y. Zhang and C. Burkhart formulated the research problem by identifying the primary challenges and applications. X. Li and Y. Zhang then developed the idea of the present work and proposed the workflow. Code implementation was done by X. Li with the help of H. Zhao and Y. Zhang on machine configuration, Linux setup and sample collection. Design of numerical experiments were developed by continuous discussion between X. Li, Y. Zhang, C. Burkhart, W. Chen and L. C. Brinson. X. Li drafted the manuscript and continuously improved the writing with the help from Y. Zhang, H. Zhao, C. Burkhart, W. Chen and L.C. Brinson. All authors reviewed the manuscript and contributed to the revision.

# Competing financial interests

The authors declare no competing financial interests.

# Data availability statement

The source code of this work will be made available upon request to the corresponding author.

**Appendix I. The plots of two-point correlation function for the material systems in Fig. 2**

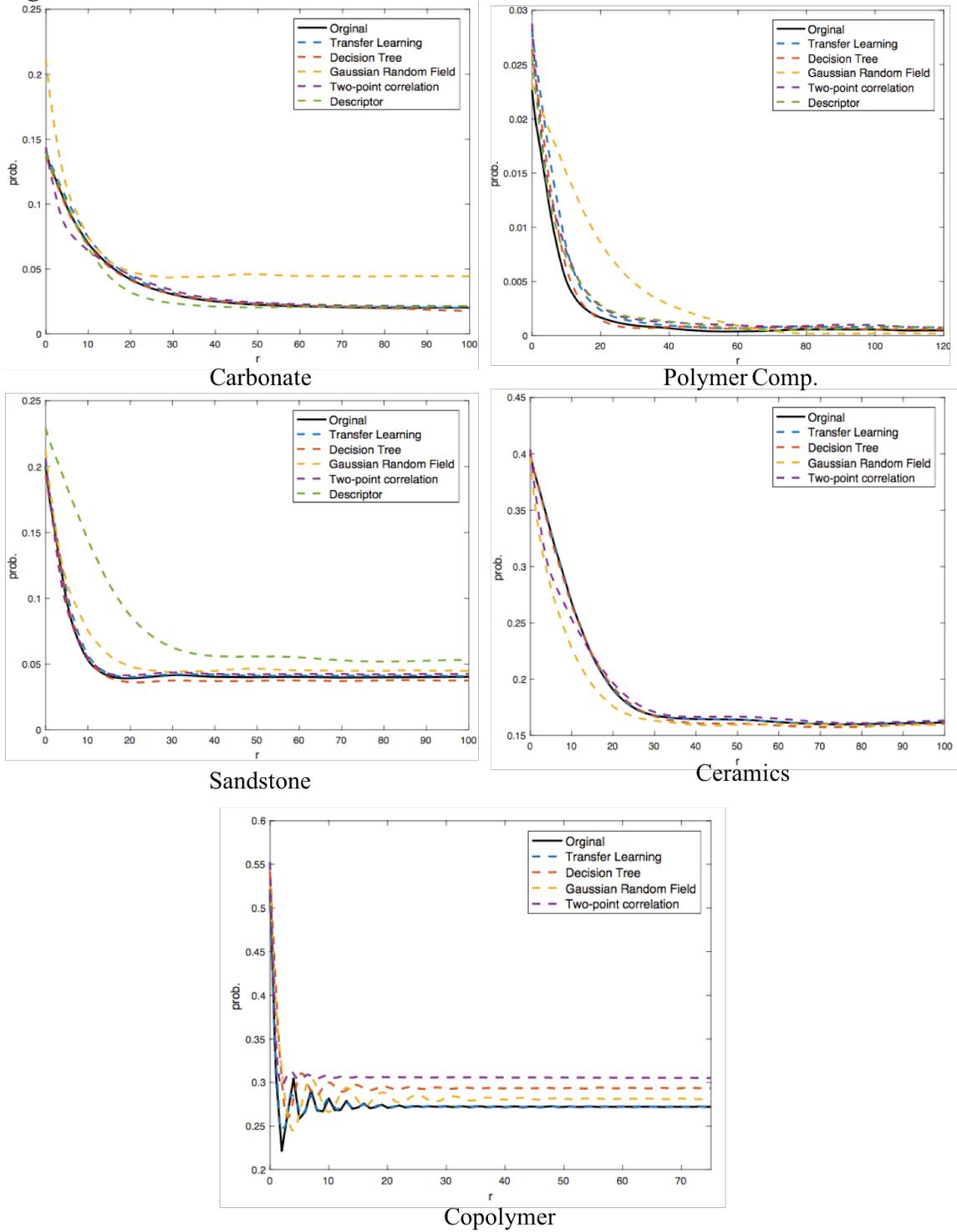

**Appendix II. The plots of lineal-path correlation function for the material systems in Fig. 2**

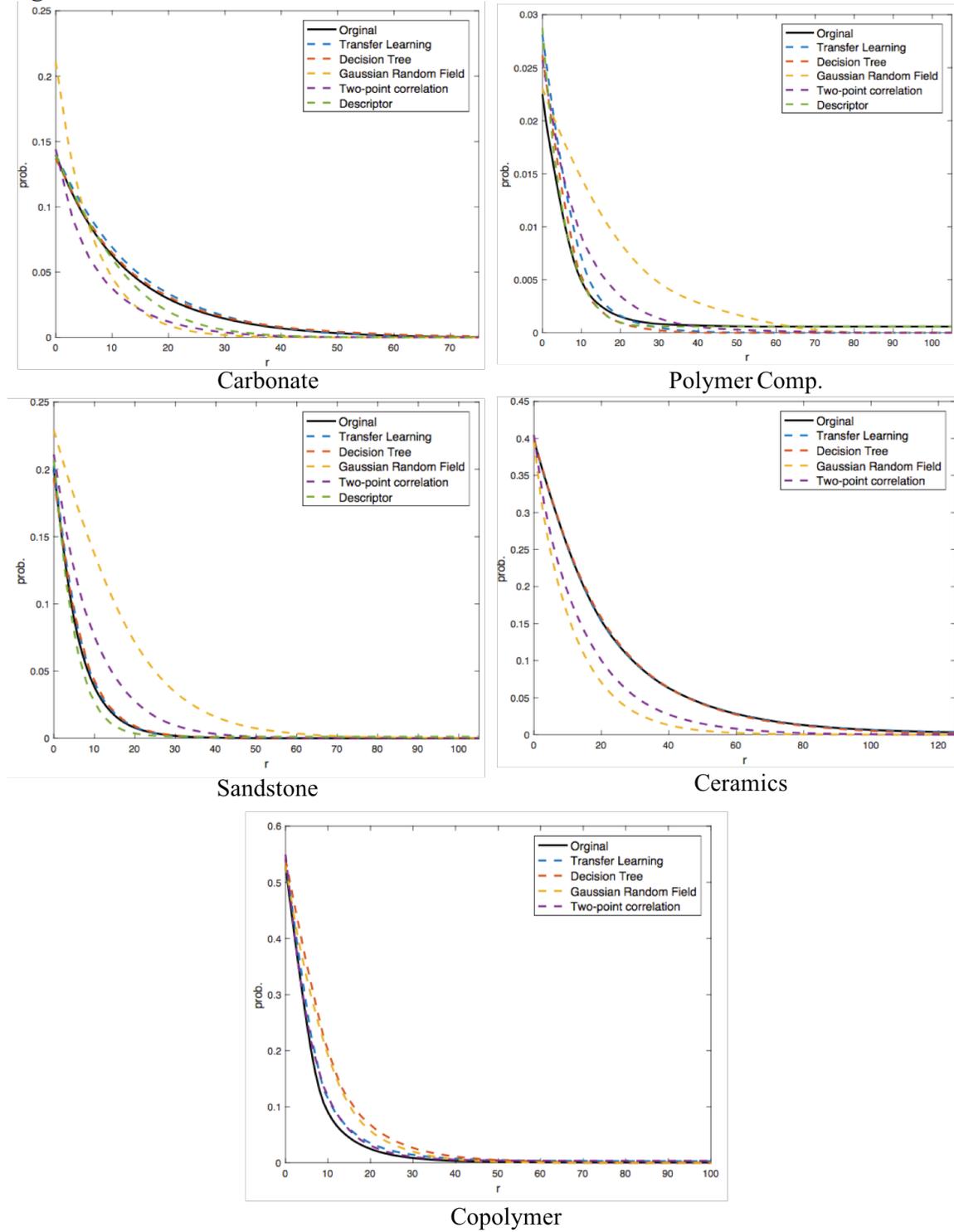